\documentclass[%
  amsmath,amssymb,
  aps,
]{revtex4-1}

\usepackage{graphicx}
\usepackage{dcolumn}
\usepackage{bm}

\begin{document}

%\preprint{}

\title{Systematic study of $\alpha$-decay half-lives of super-heavy nuclei with 106$\leq$Z$\leq$118}% Force line breaks with \\

\author{O.N.Ghodsi}
\homepage{o.nghodsi@umz.ac.ir}
\author{M.Hassanzad}

\affiliation{Department of Physics, Faculty of Basic Sciences,\\
University of Mazandaran, P.O.Box 47416-416, Babolsar, Iran\\}

\begin{abstract}

The $\alpha$- decay half-lives of the superheavy nuclei are systematically studied using different versions of proximity potential and a exact method to calculate Coulomb potential between spherical and deformed nuclei in the framework of the double folding model. To reproduce the $\alpha$-decay half-life, the experimental $\alpha$-decay energy and Wentzel-Kramers-Brillouin approximation have been used. It is found that the computed values by the Ng$\hat{o}$ 80 are in good compromise with the experimental half-lives in comparison with other versions. Also, by using this version and within $Q_{WS4}$ for determination $\alpha$-decay energies for superheavy elements, we had predicted the $\alpha$-decay half-lives for superheavy nuclei which have not been reported yet. The long half-lives with magnitude about $100$ seconds are predicted for the superheavy nuclei which are not in stability islands which indicating remarkable stability in comparison with their neighbors. These results are also in good agreement with the predictions of other semi-empirical formulas.

\begin{description}
\item[PACS numbers]

\item[Key Words]
$\alpha$-decay, Superheavy Nuclei, Proximity Potential, Half-Life
\end{description}
\end{abstract}

\maketitle

%\tableofcontents

\section{\label{sec:level1}INTRODUCTION}
The $\alpha$-radioactivity was first observed by Rutherford \cite{rutherford1908electrical} and then formulated in $1928$, independently, by Gamow \cite{gamow1928quantentheorie} and by Condon and Gurney \cite{condon1928eu} based on quantum tunneling effect. Subsequently, microscopic \cite{lovas1998rg,stewart1996alpha,delion2003anisotropic,silitextordfeminineteanu2007alpha} and macroscopic \cite{xu2006c,medeiros2006systematics,samanta2007predictions,bhagwat2008alpha} cluster and fission like \cite{duarte2002half,poenaru2006alpha} models were proposed for $\alpha$-decay.

Many investigations have been done theoretically and experimentally on the $\alpha$-decay. From the theoretical point of view, researches are applied to find an appropriate formalism for calculating the $\alpha$-decay half-life such as the generalized liquid drop model (GLDM), cluster model, density dependent M3Y (DDM3Y) effective interaction, unified fission model (UFM) \cite{xu2004c,xu2006c,buck1992alpha,buck1994alpha,xu2006mean,ni2009d,ni2009microscopic,royer2008recent,dong2008gldm,moustabchir2001analytic,royer2002formation,royer2004entrance,royer2000g,dong2010alpha2,samanta2007predictions,chowdhury2006pr,bhattacharya2008m,zhang2009determination}, and deformed proximity potential \cite{Santhosh2017even, Santhosh2018odd}. Experimental identification of the new elements using $\alpha$-decay is performed, because $\alpha$-decay is the dominant decay mode in superheavy nuclei (SHN) which synthesizing by hot, warm and cold fusion reactions \cite{hofmann2000s,oganessian2007heaviest} that the study of these nuclei is an interesting and popular subject in nuclear physics, which contributes to developing the concepts such as the stability islands, magic numbers, spin-parity and deformed nuclei.

Recently, different versions of nuclear proximity potential have been used to calculate the $\alpha$-decay half-life. The study of heavy and SHN using $14$ proximity potentials for even-even nuclei indicated the capability of this potential to estimate the $\alpha$-decay half-life \cite{yao2015comparative}. In another study, $28$ proximity potentials were used and $344$ nuclei have been investigated in the range of atomic numbers 52$\leq$Z$\leq$107 \cite{ghodsi2016systematic} in which the root mean square deviations (RDMS) for three versions were under unity. Furthermore, for even-even, even-odd, odd-even and odd-odd nuclei showed that RDMSs for even-even parent nuclei were significantly decreased.
The proximity potential is used to study the properties of the SHN $\alpha$-decay, such as $^{(256-339)}$110 \cite{santhosh2017decay}, and also for the $\alpha$-decay chains $^{(271-310)}$118 \cite{santhosh2014alpha}, $^{(270-301)}$117 \cite{santhosh2012feasibility,santhosh2012alpha}, $^{(271-294)}$115 \cite{santhosh2011alpha}, $^{255-314}$113 \cite{santhosh2016predictions}, and $^{(255-350)}$111 \cite{santhosh2017alpha} in which the half-lives estimation were in good agreement with the experimental values.

Another interesting item in $\alpha$-decay studies is the nuclear property predictions such as half-lives of the nuclei which are not in the stability islands. In these predictions the $Q_\alpha$ values play a crucial role in half-life calculations. The $Q_{\alpha}$ values can be calculated within different empirical and theoretical relationships \cite{dong2010j,dong2010alpha,moller1995nuclear,wang2014surface,bhagwat2014simple,kirson2008mutual,goriely2015further,aboussir1995nuclear,liu2006applications,duflo1995microscopic,qi2015theoretical,koura2005nuclidic,nayak1999rc}. The SHN have a very short $\alpha$-decay half-life, commonly. Hence, by finding a theoretical model that can well reproduce the $\alpha$-decay half-lives for the nuclei that their experimental data are available, we are able to predict the $\alpha$-decay half-lives for nuclei which their experimental data have not been reported yet. Therefore, in the present article, we study the $\alpha$-decay of SHN in the atomic number range $106 \leq Z \leq 118$ using the various versions of proximity formalism. The theoretical framework is introduced in Section~\ref{sec:level2}. Results and corresponding discussions are given in Section~\ref{sec:level3}. And the conclusion of entire work in Section~\ref{sec:level4}.

\section{\label{sec:level2}THEORETICAL FRAMEWORK}

\subsection{Proximity potential}

Total interaction potential V$_{T}$(r) between $\alpha$-particle and daughter nucleus is taken as follows:
\begin{equation}
{V_{T}(r)}={V_{N}(r)+V_{C}(r)+V_{l}(r)}.
\end{equation}
Here, V$_N$(r), V$_C$(r) and V$_l$(r) are the nuclear potential, Coulomb potential, and centrifugal potential, respectively. Since, the spin-parity of SHN are not known yet, to have a precise prediction we neglecting the centrifugal potential contribution in total interaction potential.
For the calculation of nuclear potential V$_N$(r), the proximity potential \cite{blocki1977proximity,myers2000wd,moller1981nuclear} is applied which was first used by Shi and Swiatecki \cite{shi1985yj} and then Gupta \cite{malik1989ss}. In recent years, many modifications and refinements have been offered over original proximity potential \cite{dutt2010comparison}. The nuclear part in proximity potential for two spherical nuclei is described as:
\begin{equation}
{V_{N}(r)}={4\pi\gamma b \overline{R}\Phi(\xi)},
\end{equation}
where $\gamma$ is the nuclear surface tension coefficient taken from Myers and Swiatecki formula \cite{myers1966anomalies} and has the following form:
\begin{equation}
{\gamma}={\gamma_{0} [1-k_{s} A_{s}^2 ]}.
\end{equation}
Here, $A_{s}=(N-Z)/(N+Z)$, $k_{s}$, and $\gamma_{0}$ are asymmetric parameters, surface asymmetry constant and surface energy constant, respectively. The width (diffuseness) of nuclear surface b is considered close to the unity. The mean curvature radius in term of S\"{u}ssmann's central radius $C_{i}$ is as follows:
\begin{equation}
{\overline{R}}=\frac{(C_{1} C_{2})}{(C_{1}+C_{2})},
\end{equation}
where
\begin{equation}
{C_{i}}={R_{i} [1-(\frac{b}{R_{i}})^2]},{(i=1,2)}.
\end{equation}
$R_{i}$ is the effective sharp radius, read as:
\begin{equation}
{R_{i}}={1.28A_{i}^\frac{1}{3}-0.76+0.8A_{i}^{-\frac{1}{3}}},{(i=1,2)}.
\end{equation}
Here, index \emph{i} refers to the $\alpha$-particle and daughter nuclei. The parametrization of dimensionless universal function $\Phi(\xi)$ is as follows:
\begin{widetext}
\begin{equation}
\Phi(\xi)= \left\{ \begin{array}{lll}
-\frac{1}{2} {(\xi-2.54)}^2-0.0852{(\xi-2.54)}^3 & \mbox{for} & \xi \leq 1.2511, \\
-3.437 exp(\frac{-\xi}{0.75})                    & \mbox{for} & \xi \geq 1.2511,
\end{array}\right.
\end{equation}
\end{widetext}
where $\xi=s/b$ is the minimum separation distance, which only depends on separation distance $s=r-C_{1}-C_{2}$ fm. This proximity model was labeled as Proximity $1977 (Prox.77)$.
Different modifications values on the surface asymmetry constant and surface energy constant that leads to different versions of Prox.77 shown in Table~\ref{tab:table1}.
Using the energy density formalism and Fermi distributions for the nuclear densities Ng$\hat{o}$ 80 and collaborators parameterized the nucleus-nucleus interaction potential in the spirit of proximity concept. The interaction potential can be divided into the geometrical factor and a universal function. The nuclear part of the parameterized potential is defined as \cite{ngo1980h}:
\begin{equation}
{V_{N}^{Ng\hat{o} 80}(r)}={\bar{R} \phi(r-C_{1}-C_{2})},
\end{equation}
where the nuclear radius $R_{i}$ reads as:
\begin{equation}
{R_{i}}={\frac{NR_{ni}+ZR_{pi}}{A_{i}}},    (i=1,2).
\end{equation}
The equivalent sharp radius for protons and neutrons are given
as:
\begin{equation}
{R_{pi}}={r_{0_{pi}} A_{i}^{1/3}}; {R_{ni}}={r_{0_{ni}}A_{i}^{1/3} },
\end{equation}
where $r_{0_{pi}}=1.128 fm$ and $r_{0_{ni}}=1.1375+1.875\times10^{-4}A_{i} fm$. The universal function $\phi(r-C_{1}-C_{2})$ is given by:
\begin{equation}
\Phi(\xi)= \left\{ \begin{array}{lll}
-33+5.4(s-s_{0})^2                                  & \mbox{for} & s <    s_{0}, \\
-33 exp(\frac{-1}{5}(s-s_{0})^2)                    & \mbox{for} & s \geq s_{0},
\end{array}\right.
\end{equation}
with $ s_{0} =-1.6$ fm. This potential labeled as Ng$\hat{o}$ 80.
The details of other versions of proximity potentials which used in this work are introduced in Ref. \cite{myers2000wd,dutt2010modified,bass1974fusion,bass1977nucleus,christensen1976evidence,reisdorf1994heavy,winther1995dissipation,denisov2002interaction,guo2013study,blocki1981generalization}. Finally, the half-life can be obtained as:
\begin{equation}
{T_{\frac{1}{2}}}={\frac{\ln2}{\nu_{0} P}}.
\end{equation}
Here, $\nu_{0}$ is the assault frequency which is related to the oscillation frequency $\omega$:
\begin{equation}
{\nu_{0}}={\frac{\omega}{2\pi}}={\frac{(2n_{r}+l+\frac{3}{2})\hbar}{(2\pi\mu R_{n}^2)}}={\frac{(G+\frac{3}{2})}{(1.2\pi\mu R_{0}^2 )}},
\end{equation}
where $R_{n}^2=\frac{3}{5} R_{0}^2$ \cite{myers2000wd} and $G=2n_{r}+l$ is the global quantum number \cite{xu2004c}:
\begin{equation}
G = 2n_{r} + 1 = \left\{ \begin{array}{lll}
22 & \mbox{for} & N >126            \\
20 & \mbox{for} & 82 < N \leq 126 . \\
18 & \mbox{for} & N\leq82
\end{array}\right.
\end{equation}
The $\alpha$-decay penetration probability $P_{\alpha}$ using the WKB semi classical approximation defined as:
\begin{equation}
{P}={exp\{{ {-\frac{2}{\hbar}} \int^{r_{b}}_{r_{a}} \sqrt{2\mu(V_{T}(r)-Q_{\alpha})}}\,dr\}},
\end{equation}
Where ${\mu}= m{\frac{A_{\alpha}+A_{d}}{A_{\alpha} A_{d}}}$ is the reduced mass which $A_{\alpha}=4$ and $A_{d}$ is daughter nucleus. The $r_{a}$ and $r_{b}$ are the turning points, which obtain from $V_{T}(r_{a} )=Q_{\alpha}=V_{T} (r_{b})$.
Realistic density distributions and the double folding model have been used to derive the Coulomb potential for spherical-deformed nuclear pair \cite{ismail2003coulomb}. In this model, the interaction Coulomb potential between spherical-deformed or deformed-deformed nuclei with separation distance $\vec{R}$ between their centers is given by:
\begin{equation}
{V_{C}(\vec{R})}={\int \int d\vec{r_1} d\vec{r_2} \frac{1}{|\vec{s}|} \rho_P (\vec{r}_1) \rho_T (\vec{r}_2)},
\end{equation}
where $\vec{S}=\vec{R}+\vec{r}_1+\vec{r}_2$. $\rho_P$ and $\rho_T$ show the nuclear charge distribution in the projectile and target nuclei which are normalized to the total charge, respectively.
Restricting our derivation to be for spherical-deformed nuclear pair with the coordinates that define as:
\begin{equation}
{G(\vec{R},\beta,s)}={\int \rho_T(\vec{R}+\vec{r}) \rho_P(\vec{r}+\vec{s})d\vec{r} },
\end{equation}
where $\beta$ is the orientation angle of the deformed nucleus. After solving and substituting into equ(13), $V_{c}(\vec{R},\beta)$ becomes:
\begin{widetext}
\begin{equation}
{V_{C}(\vec{R},\beta)}={8 \int_{0}^{\infty} \int_{0}^{\infty} s ds j_0 (ks) k^{2} dk \int d\vec{r} \rho_T(\vec{R}+\vec{r}) j_0 (kr) \int x^{2} dx j_0 (kx)  \rho_P(x) }.
\end{equation}
\end{widetext}
The the charge density distribution of the deformed nucleus is then assumed to be:
\begin{equation}
{\rho(r,\theta)}={\frac{\rho_0}{1+e^{\frac{r-R(\theta)}{a}}} },
\end{equation}
where the $R(\theta)=r_0 [1+\beta_{2} Y_{20}(\theta,0)+\beta_{4} Y_{40}(\theta,0)+...]$ is the half density radius of this Fermi distribution. $\beta_2$ and $\beta_4$ are, respectively, the quadrupole and hexadecapole deformation parameters of the residual daughter nucleuswhich their numerical values are taken from Ref.\cite{moller2016nuclear}. the parameters $r_0$ and $a$ of the density distribution are suggested at $r_0 = 1.07 A_{d} ^{\frac{1}{3}}$ and $a=0.54$ fm \cite{BohrMottelson1998}.

\subsection{Semi-empirical relationship for $\alpha$-decay}

One of the purposes of this study is to predict the half-life of the for SHN's $\alpha$-decay for which the experimental data of half-life have not been reported yet. Hence, in order to compare our obtained results with other predictions, some semi-experimental relationships which used in this work are summarized in the following.

\subsubsection{The Viola-Seaborg-Sobiczewski (VSS) semi-empirical relationship}

One of the most famous formulae for calculating alpha decay half-lives is the five parameter formula offered by Viola and Seaborg \cite{viola1966nuclear}.
\begin{equation}
{\log_{10}(T_{\frac{1}{2}})=(aZ+B) Q^{-\frac{1}{2}} + cZ + D + h_{log}},
\end{equation}
where Z is the atomic number of the parent nucleus and a, b, c and d are $1.66175$, $-8.5166$, $-0.20228$ and $-33.9069$, respectively \cite{sobiczewski1989deformed}, and
\begin{equation}
h_{log} = \left\{ \begin{array}{llll}
0     & \mbox{for} & Z=even, & N=even, \\
0.772 & \mbox{for} & Z=odd,  & N=even, \\
1.066 & \mbox{for} & Z=even, & N=odd,  \\
1.114 & \mbox{for} & Z=odd,  & N=odd.
\end{array}\right.
\end{equation}

\subsubsection{The analytical formula for $\alpha$-decay half-life}

An analytical formula for $\alpha$-decay half-lives has been developed by Royer \cite{royer2000g} and is given by
\begin{equation}
{\log_{10}(T_{\frac{1}{2}})=a+ bA^\frac{1}{6} \sqrt{Z}+\frac{cZ}{\sqrt{Q_{\alpha}}}},
\end{equation}
where A and Z represent the mass and charge number of parent nuclei. The constant a, b, and c are
\begin{widetext}
\begin{equation}
h_{log} = \left\{ \begin{array}{llllll}
a=-25.31 & b=-1.1629 & c=1.5864 & \mbox{for} & Z=even, & N=even, \\
a=-26.65 & b=-1.0859 & c=1.5848 & \mbox{for} & Z=even, & N=odd,  \\
a=-25.68 & b=-1.1423 & c=1.5920 & \mbox{for} & Z=odd,  & N=even, \\
a=-29.48 & b=-1.1130 & c=1.6971 & \mbox{for} & Z=odd,  & N=odd.
\end{array}\right.
\end{equation}
\end{widetext}

\subsubsection{The universal decay law}

New universal decay law (UDL) for $\alpha$ and cluster decay modes was introduced by Qi et al. \cite{qi2009c,qi2009c}.
\begin{equation}
{\log_{10}(T_{\frac{1}{2}})=aZ_{c} Z_{d} \sqrt{\frac{A}{Q_{c}}} +b \sqrt{AZ_{c} Z_{d} (A_{d}^\frac{1}{3} + A_{c}^\frac{1}{3})}+c},
\end{equation}
where $A=\frac{A_{c} A_{d}}{A_{c}+A_{d}}$ and the constant $a=0.4314$, $b=-0.4087$ and $c=-25.7725$ are determined by fitting to experimental of both $\alpha$ and cluster decays \cite{qi2009c}.

\section{\label{sec:level3}RESULTS AND DISCUSSION}

\begin{figure*}
\includegraphics [width=.90\textwidth,origin=c,angle=0] {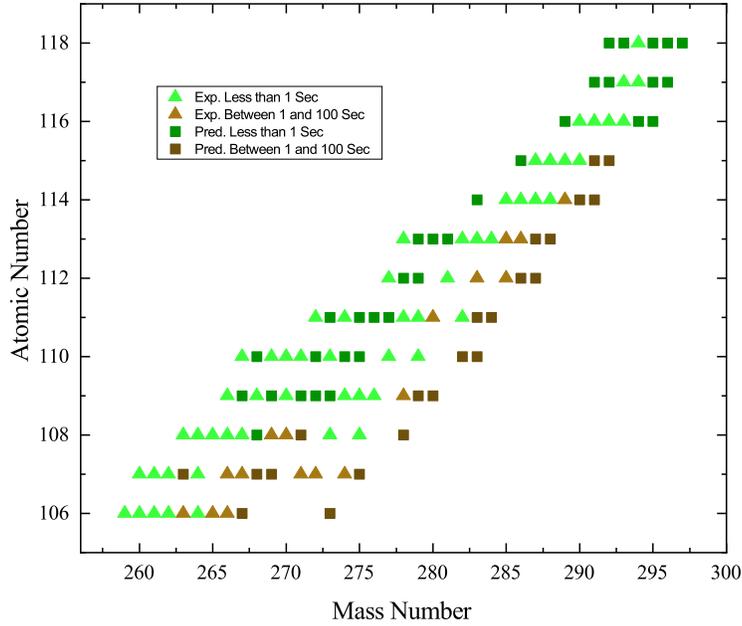} % Here is how to import EPS art
\caption{\label{fig:figure1} Predicted and experimental data for $\alpha$-decay half-lives of SHN categorized in two groups with respect to mass number.}
\end{figure*}

Since the nuclear interaction is function of some parameters such as $K_{s}$, $\gamma$ and also the distance between fragments that these parameters subsequently can effect on nuclei half-lives, we were interested to seek a version of proximity potential proceeding to reproduce experimental half-lives well. For such purpose, the $\alpha$-decay half-lives of $70$ SHN with 106$\leq$Z$\leq$118 was carried out using $28$ versions of proximity potential and the Coulomb potential for spherical-deformed nuclear pair. The barrier penetrability of the $\alpha$ particle in a deformed nucleus is different in different directions. The averaging of penetrability over different directions is done using the equation:
\begin{equation}
{P}={\frac{1}{2} \int_{0}^{\pi} P(Q,\theta) \sin(\theta)d\theta }.
\end{equation}
In order to choose the best version of proximity potential, the following standard deviation (SD) equation has been used which is defined as below:
\begin{equation}
{SD}={\sqrt{\frac{1}{N} \sum_{i=1}^N(T_{\frac{1}{2},i}^{Theo}-T_{\frac{1}{2},i}^{Exp})^2}}.
\end{equation}
Here, N is the number of $\alpha$ emitters, $T_{\frac{1}{2}}^{Theo}$ is the calculated half-life and $T_{\frac{1}{2}}^{Exp}$ is the experimental values for each nucleus which are listed in Ref. \cite{oganessian2006synthesis,oganessian2007synthesis,oganessian2005yu,oganessian2004measurements,oganessian2004experiments,wang2012ame2012}.
The standard deviation of half-life calculation which obtained using experimental $Q_{\alpha}$ are shown in Table~\ref{tab:table2}, which indicates that among the different versions of proximity potensial Ng\^{o} $80$, $\gamma$-MS $1966$, and $\gamma$-PD-LDM $2003$ have the least SD values than the other versions so can compute the $\alpha$-decay half-life in the selected region with higher accuracy. Therefore, these selected versions can be a good candidate for estimating the SHN's half-lives which are not available in stability islands.
The number of nuclei's half-lives calculations that the proximity potential versions were able to calculate them are quoted beside their standard deviations in Table~\ref{tab:table2}. Therefore, the proximity potential versions $\gamma$-MS $1967$, $\gamma$-RR $1984$, $\gamma$-PD NLD $2003$, $\gamma$-PD-LSD $2003$, and Prox.00 have been neglected in SD comparison because of the number of nuclei which are included by these versions were less than the others.
The $\alpha$-decay half-lives that derived from the most suitable versions of current calculations within experimental $\alpha$-decay energies are arranged in the seventh to ninth column of Table~\ref{tab:table3}. For quantitative analysis, the SD which are obtained within VSS, Royer, and UDL models for same nuclei which pointed out in Table \ref{tab:table3} are 67.5236, 12.5848, and 5.1898, respectively, although one particular nucleus make SD of VSS substantial. This table indicating that the Ng\^{o} $80$ is well reproduced the experimental half-lives of these nuclei.
Since the nuclear properties like $\alpha$-decay energies have not been verified for unknown SHN in stability islands yet, the various theoretical formalizations such as $Q_{WS4}$ \cite{wang2014surface} and two fitting $Q_{\alpha}$ formulas which based on the local liquid drop model \cite{dong2010alpha,dong2010j} have been employed for predicting their alpha decay energies. Researches confirm that these models predict with Z$\geq$100, well, that the $Q_{WS4}$ model is in a better agreement with experimental data of $Q_{\alpha}$ values in comparison with the others \cite{wang2015systematic}.
Consequently, the nuclear potentials are described by Ng\^{o} 80 and $\alpha$-decay energies are obtained from $Q_{WS4}$ were used to calculate the $\alpha$-decay half-lives for SHN in the specified region which have not been existed in the stability island yet. The results of these predictions are compared with the half-lives of the semi-empirical relationship VSS and the Royer analytical formula and UDL which are listed in Table~\ref{tab:table4}. As can be seen from Table~\ref{tab:table4}, our predictions are compatible with the prediction of other models.
To find a better perspective, we had classified the nuclei with available experimental data for the $\alpha$-emitter system and also nuclei which predicted so far in two groups, one with half-lives below $1$ second and the other with half-lives between $1$ to $100$ seconds which are presented in Figure~\ref{fig:figure1}.

\section{\label{sec:level4}CONCLUSION}

The estimated half-lives of the $\alpha$-decay for $70$ SHN in the atomic number range 106$\leq$Z$\leq$118 is estimated using $28$ different versions of the proximity potentials and also a deformed method for Coulomb potential. In these calculations, experimental $\alpha$-decay energies have been used to obtain the penetration probabilities within the WKB approximation. To choose the best proximity potential versions we had used standard deviation between the calculated and experimental half-lives. The results have shown that Ng\^{o} $80$ is the best choice in this region. The Ng\^{o} $80$ and $Q_{WS4}$ have been used to predict the unknown SHN's half-lives. Furthermore, by increasing neutron number in the regions in which the most SHN have half-lives with less than one second, it is predicted that some more stable unknown SHN with the half-lives by magnitude about $100$ seconds can be expected to be there in that regions. Also, we have confirmed that our half-lives were in good compromise with the semi-empirical Viola-Seaborg-Sobiczewski relationship, Universal Decay Law and Royer analytical formulas. In further work we are going to employ the deformed proximity formalism, in order to increase the accuracy of half-lives calculations.

\newpage
\begin{ruledtabular}
\begin{table*}
\caption{\label{tab:table1}Different values for surface asymmetric constant and surface energy constant which determine different versions of prox. 77. }
\begin{tabular}{llll}
$\gamma$-set          &$\gamma_{0}$ \scriptsize\emph{$[MeV/fm^2]$}& $K_{s}$ & References                   \\
\hline
$\gamma$-MS 1967      & 0.9517                                    & 1.7826  & \cite{myers1966anomalies}    \\

$\gamma$-Ms 1966      & 1.01734                                   & 1.79    & \cite{myers1966nuclear}      \\

$\gamma$-MN 1976      & 1.460734                                  & 4.0     & \cite{moller1976macroscopic} \\

$\gamma$-KNS 1979     & 1.2402                                    & 3.0     & \cite{krappe1979hj}          \\

$\gamma$-MN-{I} 1981  & 1.1754                                    & 2.2     & \cite{moller1981nuclear}     \\

$\gamma$-MN-{II} 1981 & 1.27326                                   & 2.5     & \cite{moller1981nuclear}     \\

$\gamma$-MN-{III} 1981& 1.2502                                    & 2.4     & \cite{moller1981nuclear}     \\

$\gamma$-RR 1984      & 0.9517                                    & 2.6     & \cite{royer1984fission}      \\

$\gamma$-MN 1988      & 1.2496                                    & 2.3     & \cite{moller1988p}           \\

$\gamma$-MP 1988      & 1.65                                      & 2.3     & \cite{kumar2012systematic}   \\

$\gamma$-MN 1995      & 1.25284                                   & 2.345   & \cite{moller1995nuclear}     \\

$\gamma$-PD-LDM 2003  & 1.08948                                   & 1.9830  & \cite{pomorski2003nuclear}   \\

$\gamma$-PD NLD 2003  & 0.9180                                    & 0.7546  & \cite{pomorski2003nuclear}   \\

$\gamma$-PD LSD 2003  & 0.911445                                  & 2.2938  & \cite{pomorski2003nuclear}   \\
\end{tabular}
\end{table*}
\end{ruledtabular}

\newpage
\begin{ruledtabular}
\begin{table*}
\caption{\label{tab:table2} SDs for $\alpha$-decay half-lives of different versions of the proximity potential.  data set consists of 70 parent nuclei. The numbers in parentheses are the number of nuclei under calculation.}
\begin{tabular}{llllll}
Proximity model         & SD             &Proximity model          & SD            &Proximity model     & SD           \\
\hline
$\gamma$-MS 1967        & 4.7923 (45)    & $\gamma$-MN 1995        & 6.2595 (70)   & Bass 80            & 6.3370 (70) \\

$\gamma$-MS 1966        & 4.7238 (70)    & $\gamma$-PD-LDM 2003    & 5.7812 (70)   & CW 76              & 6.5296 (70) \\

$\gamma$-MN 1976        & 6.3570 (70)    & $\gamma$-PD-NLD 2003    & 5.4113 (49)   & BW 91              & 6.2794 (70) \\

$\gamma$-KNS 1979       & 6.1472 (70)    & $\gamma$-PD-LSD 2003    & 0.0001 (5)    & AW 95              & 6.4893 (70) \\

$\gamma$-MN-{I} 1981    & 6.1025 (70)    & Prox. 00                & 5.4590 (60)   & Ng\^{o} 80         & 4.6053 (70)  \\

$\gamma$-MN-{II} 1981   & 6.2713 (70)    & Prox. 00DP              & 6.5516 (70)   & Denisov            & 6.5569 (70) \\

$\gamma$-MN-{III} 1981  & 6.2459 (70)    & Prox. 2010              & 6.5527 (70)   & Denisov DP         & 6.5573 (70)  \\

$\gamma$-RR 1984        & 0.1056 (27)    & Dutt 2011               & 6.0652 (70)   & Guo 2013           & 6.4813 (70)  \\

$\gamma$-MN 1988        & 6.2555 (70)    & Bass 73                 & 6.2916 (70)   &                    &              \\

$\gamma$-MP 1988        & 6.4915 (70)    & Bass 77                 & 9.2451 (70)   &                    &              \\
\end{tabular}
\end{table*}
\end{ruledtabular}

\newpage
\begin{ruledtabular}
\begin{table*}
\caption{\label{tab:table3} Comparisons between the experimental and theoretical $\alpha$-decay half-lives of SHN in the unit of sec.}
\begin{tabular}{lllllllll}
${A}$ & ${Z}$ & ${\beta_{2}}$ & $\beta_{4}$ & $Q_\alpha^{Exp}$ \scriptsize\emph{[MeV]}  & ${T_{\frac{1}{2}, \alpha}^{Exp}}$ & ${T_{\frac{1}{2}, \alpha}^{Ng{\hat{o}}80}}$ & ${T_{\frac{1}{2}, \alpha} ^{MS 1966}}$ & ${T_{\frac{1}{2}, \alpha}^{PD-LDM 2003}}$ \\
\hline
294   &  118   &   0.064   &   -0.022   &   11.810   &   8.90E-04   &   5.93E-04   &    1.87E-04   &    1.05E-04 \\
294   &  117   &   0.053   &   -0.011   &   11.070   &   5.10E-02   &   1.79E-02   &    6.30E-03   &    3.31E-03 \\
293   &  117   &   0.064   &   -0.010   &   11.180   &   1.40E-02   &   9.94E-03   &    3.34E-03   &    1.78E-03 \\
293   &  116   &  -0.021   &    0.012   &   10.670   &   5.30E-02   &  	9.58E-02   &  	3.70E-02   &  	1.86E-02 \\
292   &  116   &  -0.021   &    0.012   &   10.800   &   1.80E-02   &  	4.54E-02   &  	1.65E-02   &  	8.50E-03 \\
291   &  116   &   0.064   &   -0.010   &   10.890   &   1.80E-02   &  	2.62E-02   &  	9.08E-03   &  	4.78E-03 \\
290   &  116   &   0.064   &   -0.010   &   11.000   &   7.10E-03   &  	1.44E-02   &  	4.84E-03   &  	2.59E-03 \\
290   &  115   &   0.075   &   -0.022   &   10.400   &   1.60E-02   &  	2.48E-01   &  	1.15E-01   &  	4.74E-02 \\
289   &  115   &   0.075   &   -0.010   &   10.520   &   2.20E-01   &  	1.24E-01   &  	4.98E-02   &  	2.33E-02 \\
288   &  115   &   0.075   &   -0.010   &   10.610   &   8.70E-02   &  	7.44E-02   &  	2.70E-02   &  	1.36E-02 \\
287   &  115   &   0.075   &    0.002   &   10.740   &   3.20E-02   &  	3.43E-02   &  	1.16E-02   &  	6.12E-03 \\
289   &  114   &   0.086   &   -0.033   &   9.960    &   2.70E+00   &  	2.02E+00   &  	1.13E+00   &  	4.08E-01 \\
288   &  114   &   0.086   &   -0.021   &   10.090   &   8.00E-01   &  	8.70E-01   &  	4.10E-01   &  	1.69E-01 \\
287   &  114   &   0.086   &   -0.009   &   10.160   &   4.80E-01   &  	5.92E-01   &  	2.63E-01   &  	1.11E-01 \\
286   &  114   &   0.086   &   -0.009   &   10.330   &   1.30E-01   &  	2.03E-01   &  	7.90E-02   &  	3.74E-02 \\
285   &  114   &   0.086   &    0.003   &   10.520   &   1.50E-01   &  	6.22E-02   &  	2.19E-02   &  	1.12E-02 \\
286   &  113   &   0.108   &   -0.032   &   9.770    &   2.00E+01   &  	8.92E+00   &  	6.81E+00   &  	1.75E+00 \\
285   &  113   &   0.130   &   -0.042   &   10.030   &   5.50E+00   &   5.86E-01   &    2.45E-01   &   	1.13E-01 \\
284   &  113   &   0.130   &   -0.030   &   10.150   &   4.80E-01   &   2.73E-01   &    1.04E-01   &   	5.15E-02 \\
283   &  113   &   0.164   &   -0.063   &   10.260   &   1.00E-01   &   1.35E-01   &    4.96E-02   &   	2.52E-02 \\
282   &  113   &   0.175   &   -0.062   &   10.830   &   7.30E-02   &   4.02E-03   &   	1.29E-03   &   	7.39E-04 \\
278   &  113   &   0.222   &   -0.093   &   11.850   &   1.40E-03   &   1.59E-05   &   	4.45E-06   &   	2.79E-06 \\
285   &  112   &   0.130   &   -0.042   &   9.290    &   3.40E+01   &   4.44E+01   &   	2.75E+01   &   	9.16E+00 \\
283   &  112   &   0.130   &   -0.042   &   9.670    &   3.80E+00   &   3.08E+00   &   	1.39E+00   &   	6.02E-01 \\
281   &  112   &   0.198   &   -0.084   &   10.460   &   1.30E-01   &   1.61E-02   &   	5.45E-03   &   	3.07E-03 \\
277   &  112   &   0.221   &   -0.080   &   11.620   &   6.90E-04   &   2.54E-05   &   	7.34E-06   &   	4.60E-06 \\
282   &  111   &   0.153   &   -0.052   &   9.510    &   5.00E-01   &   3.86E+00   &   	1.73E+00   &   	8.00E-01 \\
280   &  111   &   0.198   &   -0.084   &   9.870    &   3.60E+00   &   3.20E-01   &   	1.19E-01   &   	6.35E-02 \\
279   &  111   &   0.209   &   -0.083   &   10.520   &   1.70E-01   &   5.05E-03   &   	1.68E-03   &   	9.91E-04 \\
278   &  111   &   0.222   &   -0.093   &   10.890   &   4.20E-03   &   5.81E-04   &   	1.83E-04   &   	1.14E-04 \\
274   &  111   &   0.232   &   -0.066   &   11.480   &   1.20E-02   &   2.70E-05   &   	7.74E-06   &   	4.88E-06 \\
272   &  111   &   0.232   &   -0.065   &   11.197   &   3.80E-03   &   2.70E-05   &   	7.74E-06   &   	4.88E-06 \\
279   &  110   &   0.198   &   -0.084   &   9.840    &   2.00E-01   &   1.68E-01   &    6.30E-02   &   	3.50E-02 \\
277   &  110   &   0.221   &   -0.081   &   10.840   &   4.10E-03   &   3.58E-04   &    1.16E-04   &   	7.27E-05 \\
273   &  110   &   0.232   &   -0.065   &   11.370   &   1.70E-04   &   2.29E-05   &   	6.73E-06   &   	4.29E-06 \\
271   &  110   &   0.232   &   -0.065   &   10.870   &   1.61E-03   &   3.99E-04   &   	1.14E-04   &   	7.00E-05 \\
270   &  110   &   0.232   &   -0.052   &   11.117   &   1.00E-04   &   1.07E-04   &   	2.97E-05   &   	1.86E-05 \\
269   &  110   &   0.243   &   -0.050   &   11.509   &   1.79E-04   &   1.32E-05   &   	3.59E-06   &   	2.30E-06 \\
267   &  110   &   0.242   &   -0.038   &   11.780   &   2.80E-06   &   3.60E-06   &   	9.45E-07   &   	6.10E-07 \\
278   &  109   &   0.198   &   -0.071   &   9.460    &   7.60E+00   &   9.75E-01   &   	3.97E-01   &    2.09E-01 \\
276   &  109   &   0.221   &   -0.081   &   9.850    &   7.20E-01   &   7.10E-02   &   	2.55E-02   &   	1.47E-02 \\
275   &  109   &   0.221   &   -0.080   &   10.480   &   9.70E-03   &   1.39E-03   &   	4.62E-04   &   	2.85E-04 \\
274   &  109   &   0.222   &   -0.079   &   9.950    &   4.40E-01   &   4.12E-02   &   	1.40E-02   &   	8.13E-03 \\
270   &  109   &   0.232   &   -0.052   &   10.180   &   5.00E-03   &   1.15E-02   &   	3.52E-03   &   	2.09E-03 \\
268   &  109   &   0.243   &   -0.050   &   10.670   &   2.10E-02   &   6.35E-04   &   	1.80E-04   &   	1.11E-04 \\
266   &  109   &   0.243   &   -0.037   &   10.996   &   1.70E-03   &   1.10E-04   &   	2.98E-05   &   	1.87E-05 \\
275   &  108   &   0.221   &   -0.080   &   9.440    &   1.90E-01   &   4.92E-01   &   	1.85E-01   &   	1.06E-01 \\
273   &  108   &   0.232   &   -0.065   &   9.730    &   7.60E-01   &   7.42E-02   &   	2.62E-02   &   	1.52E-02 \\
270   &  108   &   0.232   &   -0.052   &   9.050    &   3.60E+00   &   1.09E+01   &   	3.84E+00   &   	2.05E+00 \\
269   &  108   &   0.232   &   -0.052   &   9.370    &   9.70E+00   &   1.11E+00   &   	3.65E-01   &   	2.05E-01 \\
267   &  108   &   0.242   &   -0.038   &   10.037   &   6.50E-02   &   1.35E-02   &   	4.06E-03   &   	2.43E-03 \\
266   &  108   &   0.243   &   -0.037   &   10.346   &   2.30E-03   &   2.07E-03   &   	5.98E-04   &   	3.72E-04 \\
265   &  108   &   0.242   &   -0.025   &   10.470   &   2.00E-03   &   1.01E-03   &   	2.85E-04   &   	1.77E-04 \\
264   &  108   &   0.242   &   -0.024   &   10.591   &   1.60E-03   &   5.30E-04   &   	1.45E-04   &   	9.11E-05 \\
263   &  108   &   0.253   &   -0.022   &   10.730   &   7.40E-04   &   2.46E-04   &   	6.61E-05   &   	4.17E-05 \\
\end{tabular}
\end{table*}
\end{ruledtabular}

\newpage
\begin{ruledtabular}
\begin{table*}
\caption{\label{tab:countinue table3}  \emph{continued Table \ref{tab:table3}}.Comparisons between the experimental and theoretical $\alpha$-decay half-lives of SHN in the unit of sec.}
\begin{tabular}{lllllllll}
${A}$ & ${Z}$ & ${\beta_{2}}$ & $\beta_{4}$ & $Q_\alpha^{Exp}$ \scriptsize\emph{[MeV]}  & ${T_{\frac{1}{2}, \alpha}^{Exp}}$ & ${T_{\frac{1}{2}, \alpha}^{Ng{\hat{o}}80}}$ & ${T_{\frac{1}{2}, \alpha} ^{MS 1966}}$ & ${T_{\frac{1}{2}, \alpha}^{PD-LDM 2003}}$ \\
\hline
272   &  107   &   0.232   &   -0.065   &   9.150    &   9.80E+00   &   1.79E+00   &   	6.98E-01   &   	3.79E-01 \\
271   &  107   &   0.232   &   -0.065   &   9.490    &   1.20E+00   &   1.64E-01   &   	5.77E-02   &   	3.41E-02 \\
267   &  107   &   0.231   &   -0.040   &   9.230    &   1.70E+01   &   1.34E+00   &    4.50E-01   &   	2.52E-01 \\
266   &  107   &   0.242   &   -0.038   &   9.430    &   1.70E+00   &   3.29E-01   &    1.05E-01   &   	6.02E-02 \\
264   &  107   &   0.242   &   -0.025   &   9.960    &   4.40E-01   &   1.07E-02   &   	3.12E-03   &   	1.89E-03 \\
262   &  107   &   0.252   &   -0.010   &   10.319   &   1.02E-01   &   1.27E-03   &   	3.52E-04   &   	2.19E-04 \\
261   &  107   &   0.252   &   -0.010   &   10.500   &   1.26E-02   &   4.46E-04   &   	1.20E-04   &   	7.62E-05 \\
260   &  107   &   0.252   &    0.002   &   10.400   &   3.50E-02   &   8.71E-04   &   	2.33E-04   &   	1.45E-04 \\
265   &  106   &   0.242   &   -0.025   &   9.050    &   2.88E+01   &   1.99E+00   &   	6.58E-01   &   	3.82E-01 \\
264   &  106   &   0.242   &   -0.025   &   9.210    &   1.03E-01   &   6.74E-01   &   	2.17E-01   &   	1.26E-01 \\
263   &  106   &   0.242   &   -0.025   &   9.400    &   1.43E+00   &   1.91E-01   &   	6.05E-02   &   	3.55E-02 \\
262   &  106   &   0.252   &   -0.010   &   9.600    &   3.14E-02   &   5.00E-02   &   	1.49E-02   &   	8.96E-03 \\
261   &  106   &   0.252   &   -0.010   &   9.714    &   2.30E-01   &   5.00E-02   &   	1.49E-02   &   	8.96E-03 \\
260   &  106   &   0.252   &    0.002   &   9.901    &   7.20E-03   &   7.91E-03   &   	2.24E-03   &   	1.37E-03 \\
259   &  106   &   0.252   &    0.002   &   9.804    &   3.22E-01   &   1.56E-02   &   	4.35E-03   &   	2.65E-03 \\
\end{tabular}
\end{table*}
\end{ruledtabular}

\newpage
\begin{ruledtabular}
\begin{table*}
\caption{\label{tab:table4}  Predicted $\alpha$-decay half-lives of SHN by using the Ng\^{o} 80 in the unit of sec.}
\begin{tabular}{lllllllll}
${A}$ & ${Z}$ & ${\beta_{2}}$ & $\beta_{4}$ & $Q_\alpha^{WS4}$ \scriptsize\emph{[MeV]}  &  ${T_{\frac{1}{2}, \alpha}^{Ng{\hat{o}}80}}$ & $T_{\frac{1}{2}, \alpha}^{VSS}$ & $T_{\frac{1}{2}, \alpha}^{Royer}$ & $T_{\frac{1}{2}, \alpha}^{UDL}$ \\
\hline
297   &  118   &  -0.042   &   0.001   &  12.078   &  1.22E-04   &  1.83E-03   &  4.90E-04   &  9.14E-05 \\
296   &  118   &  -0.063   &   0.002   &  11.726   &  8.54E-04   &  1.00E-03   &  5.56E-04   &  6.75E-04 \\
295   &  118   &  -0.084   &  -0.009   &  11.876   &  3.84E-04   &  5.23E-03   &  1.52E-03   &  3.00E-04 \\
293   &  118   &   0.075   &  -0.046   &  12.214   &  7.31E-05   &  9.13E-04   &  2.88E-04   &  5.07E-05 \\
292   &  118   &   0.075   &  -0.034   &  12.212   &  7.56E-05   &  7.92E-05   &  5.25E-05   &  5.32E-05 \\
296   &  117   &  -0.032   &   0.000   &  11.477   &  1.52E-03   &  2.61E-02   &  1.13E-02   &  1.27E-03 \\
295   &  117   &  -0.052   &   0.001   &  11.270   &  5.23E-03   &  3.77E-02   &  8.40E-03   &  4.47E-03 \\
292   &  117   &   0.075   &  -0.034   &  11.724   &  4.53E-04   &  6.85E-03   &  3.17E-03   &  3.55E-04 \\
291   &  117   &   0.075   &  -0.034   &  11.690   &  5.86E-04   &  3.74E-03   &  9.80E-04   &  4.46E-04 \\
295   &  116   &  -0.021   &   0.000   &  10.748   &  5.29E-02   &  7.84E-01   &  1.78E-01   &  4.90E-02 \\
294   &  116   &  -0.042   &   0.001   &  10.639   &  1.08E-01   &  1.31E-01   &  6.48E-02   &  1.02E-01 \\
289   &  116   &   0.075   &  -0.010   &  11.146   &  6.47E-03   &  7.62E-02   &  2.20E-02   &  5.15E-03 \\
292   &  115   &  -0.042   &   0.001   &  9.906    &  6.51E+00   &  9.05E+01   &  6.12E+01   &  6.73E+00 \\
291   &  115   &  -0.042   &   0.001   &  10.166   &  1.16E+00   &  7.38E+00   &  1.61E+00   &  1.13E+00 \\
286   &  115   &   0.075   &   0.002   &  10.469   &  2.04E-01   &  2.37E+00   &  1.59E+00   &  1.76E-01 \\
291   &  114   &   0.011   &   0.000   &  9.245    &  3.61E+02   &  4.00E+03   &  8.37E+02   &  3.95E+02 \\
290   &  114   &  -0.011   &   0.000   &  9.495    &  5.63E+01   &  5.59E+01   &  2.73E+01   &  5.96E+01 \\
283   &  114   &   0.064   &   0.014   &  10.843   &  9.94E-03   &  1.10E-01   &  3.22E-02   &  7.70E-03 \\
288   &  113   &   0.054   &  -0.023   &  9.321    &  8.58E+01   &  1.16E+03   &  8.50E+02   &  9.54E+01 \\
287   &  113   &   0.064   &  -0.022   &  9.319    &  9.16E+01   &  5.36E+02   &  1.16E+02   &  1.00E+02 \\
281   &  113   &   0.064   &   0.014   &  11.247   &  4.54E-04   &  2.88E-03   &  7.67E-04   &  3.21E-04 \\
280   &  113   &   0.053   &   0.013   &  11.771   &  2.79E-05   &  3.97E-04   &  1.43E-04   &  1.76E-05 \\
279   &  113   &   0.011   &   0.000   &  12.098   &  5.67E-06   &  3.51E-05   &  1.00E-05   &  3.22E-06 \\
287   &  112   &   0.075   &  -0.046   &  9.040    &  1.84E+01   &  3.74E+03   &  7.41E+02   &  3.48E+02 \\
286   &  112   &   0.075   &  -0.034   &  9.014    &  3.80E+02   &  3.90E+02   &  1.90E+02   &  4.45E+02 \\
279   &  112   &   0.175   &  -0.062   &  11.384   &  8.79E-05   &  1.38E-03   &  3.81E-04   &  6.98E-05 \\
278   &  112   &   0.187   &  -0.072   &  11.739   &  1.28E-05   &  1.88E-05   &  1.28E-05   &  1.02E-05 \\
284   &  111   &   0.097   &  -0.032   &  8.664    &  2.60E+03   &  3.36E+04   &  2.85E+04   &  3.23E+03 \\
283   &  111   &   0.097   &  -0.032   &  9.002    &  1.77E+02   &  1.13E+03   &  2.40E+02   &  2.10E+02 \\
277   &  111   &   0.198   &  -0.084   &  11.340   &  5.28E-05   &  4.55E-04   &  1.17E-04   &  4.28E-05 \\
276   &  111   &   0.198   &  -0.084   &  11.465   &  2.77E-05   &  5.18E-04   &  1.71E-04   &  2.21E-05 \\
275   &  111   &   0.222   &  -0.093   &  11.393   &  4.20E-05   &  3.44E-04   &  9.63E-05   &  3.43E-05 \\
273   &  111   &   0.221   &  -0.080   &  11.148   &  1.80E-04   &  1.28E-03   &  3.92E-04   &  1.49E-04 \\
283   &  110   &   0.108   &  -0.044   &  8.146    &  8.69E+04   &  9.32E+05   &  1.77E+05   &  1.17E+05 \\
282   &  110   &   0.130   &  -0.043   &  8.515    &  3.48E+03   &  3.68E+03   &  1.81E+03   &  4.60E+03 \\
275   &  110   &   0.222   &  -0.093   &  10.933   &  2.33E-04   &  4.12E-03   &  1.07E-03   &  2.11E-04 \\
274   &  110   &   0.222   &  -0.093   &  10.896   &  3.05E-04   &  4.35E-04   &  2.96E-04   &  2.73E-04 \\
272   &  110   &   0.221   &  -0.080   &  10.510   &  3.32E-03   &  3.98E-03   &  2.96E-03   &  3.08E-03 \\
268   &  110   &   0.232   &  -0.065   &  11.725   &  4.79E-06   &  5.47E-06   &  4.82E-06   &  3.29E-06 \\
280   &  109   &   0.130   &  -0.043   &  8.689    &  3.35E+02   &  5.22E+03   &  3.58E+03   &  4.29E+02 \\
279   &  109   &   0.130   &  -0.042   &  9.113	   &  1.30E+01   &  9.93E+01   &  2.08E+01   &  1.53E+01 \\
273   &  109   &   0.221   &  -0.080   &  10.194   &  9.23E-03   &  7.59E-02   &  1.97E-02   &  9.34E-03 \\
272   &  109   &   0.221   &  -0.080   &  9.972	   &  4.05E-02   &  6.62E-01   &  3.32E-01   &  4.19E-02 \\
271   &  109   &   0.221   &  -0.080   &  9.789	   &  1.45E-01   &  9.71E-01   &  2.78E-01   &  1.51E-01 \\
269   &  109   &   0.232   &  -0.065   &  10.438   &  2.48E-03   &  1.76E-02   &  5.37E-03   &  2.29E-03 \\
267   &  109   &   0.232   &  -0.065   &  11.027   &  9.07E-05   &  6.28E-04   &  2.06E-04   &  7.20E-05 \\
278   &  108   &   0.130   &  -0.042   &  8.760	   &  7.88E+01   &  1.01E+02   &  5.05E+01   &  9.79E+01 \\
271   &  108   &   0.221   &   0.080   &  9.346	   &  1.23E+00   &  1.71E+01   &  4.26E+00   &  1.40E+00 \\
268   &  108   &   0.232   &  -0.065   &  9.736	   &  9.44E-02   &  1.08E-01   &  8.19E-02   &  9.87E-02 \\
275   &  107   &   0.175   &  -0.062   &  8.541	   &  1.78E+02   &  1.41E+03   &  2.99E+02   &  2.47E+02 \\
269   &  107   &   0.232   &  -0.065   &  8.605	   &  1.36E+02   &  8.55E+02   &  2.34E+02   &  1.81E+02 \\
268   &  107   &   0.232   &  -0.065   &  8.824	   &  2.53E+01   &  3.58E+02   &  2.61E+02   &  3.22E+01 \\
263   &  107   &   0.242   &  -0.038   &  8.992	   &  9.71E+00   &  4.75E+01   &  1.65E+01   &  1.05E+01 \\
273   &  106   &   0.198   &  -0.071   &  8.232	   &  9.05E+02   &  1.39E+04   &  2.66E+03   &  1.35E+03 \\
267   &  106   &   0.232   &  -0.065   &  8.274	   &  8.71E+02   &  9.85E+03   &  2.41E+03   &  1.17E+03 \\
\end{tabular}
\end{table*}
\end{ruledtabular}

\newpage

\end{document}